\documentclass[a4paper,11pt]{article}
\usepackage[margin=1.14in]{geometry}
\usepackage{tabularx}
\usepackage{multirow}
\usepackage{arydshln}
\usepackage{pdfpages}
\usepackage{float}
\makeatletter \renewcommand{\@dotsep}{10000} \makeatother
\usepackage{wrapfig}
\usepackage{authblk}
\usepackage{lmodern}
\usepackage{amsmath, amssymb, amsthm}
\usepackage{mathtools}
\usepackage{bm}
\usepackage{booktabs}
\usepackage{multirow}
\usepackage{array}
\usepackage{subcaption}
\usepackage{hyperref}
\usepackage{enumitem}
\usepackage{microtype}
\usepackage{parskip}
\usepackage{fancyhdr}
\usepackage{titlesec}
\usepackage{abstract}
\usepackage[colorinlistoftodos, textwidth=2.42cm, size=scriptsize]{todonotes}

\newcommand{\vx}{\mathbf{x}}

\newcommand{\ve}{\mathbf{e}}

\newcommand{\Energy}{E_{\theta}}

\definecolor{darkred}{rgb}{0.6, 0, 0}
\hypersetup{
	unicode,
	colorlinks,
	breaklinks,
	urlcolor=darkred,
        linkcolor=black,
        citecolor=darkred,
}

\newcommand{\be}{\begin{eqnarray}}
\newcommand{\ee}{\end{eqnarray}}
\def\be{\begin{equation}}

\def\ee{\end{equation}}
\def\bea{\begin{eqnarray}}
\def\eea{\end{eqnarray}}

\newcommand{\gsim}{\;\raisebox{-0.9ex}{$\textstyle\stackrel{\textstyle >}{\sim}$}\;}
\newcommand{\lsim}{\;\raisebox{-0.9ex}{$\textstyle\stackrel{\textstyle<}{\sim}$}\;}
\def\lsim{\raise0.3ex\hbox{$\;<$\kern-0.75em\raise-1.1ex\hbox{$\sim\;$}}}
\def\gsim{\raise0.3ex\hbox{$\;>$\kern-0.75em\raise-1.1ex\hbox{$\sim\;$}}}

\usepackage{epsfig}
\usepackage{slashed}
\usepackage[utf8]{inputenc}
\usepackage{changepage}
\usepackage{dcolumn}

\usepackage[dvipsnames]{xcolor}

\usepackage[sort&compress,numbers,square]{natbib}
\theoremstyle{plain}

\theoremstyle{definition}

\usepackage{graphicx}

\usepackage{graphicx}

\title{\bf Probabilistic Energy-Based Neural Networks for Uncertainty-Aware Collider Data Analyses} 

\vspace{8mm}
\author[1]{\Large Raymundo~Ramos\footnote{raramos@kias.re.kr }}
\author[2]{\Large Ahmed~Hammad\footnote{hammad@kias.re.kr}}
\author[3]{\\\Large Amit~Chakraborty\footnote{amit.c@srmap.edu.in }}
\author[4]{\Large Pyungwon~Ko\footnote{pko@kias.re.kr }}
\author[5,6]{\Large Stefano~Moretti\footnote{S.Moretti@soton.ac.uk , stefano.moretti@physics.uu.se }}
\affil[1]{\small Quantum Universe Center, KIAS, Seoul 02455, Korea}
\affil[2]{\small  Center of AI and Natural science, KIAS, 02455 Seoul, Republic of Korea}
\affil[3]{\small Department of Physics, SRM University-AP, Amaravati, Mangalagiri 522240, India}
\affil[4]{\small School of Physics, KIAS, Seoul 02455, Korea}
\affil[5]{\small School of Physics \& Astronomy, University of Southampton, Southampton SO17 1BJ, UK}
\affil[6]{\small Department of Physics \& Astronomy, Uppsala University, Box 516, 75120 Uppsala, Sweden}

\date{}

\begin{document}
	\maketitle
	\vspace{4mm}
	\begin{abstract}
We introduce a probabilistic  energy-based Neural Network (NN) for uncertainty-aware classification in  collider analyses. Standard discriminative classifiers can give highly confident predictions even for events that are different from the data used during training. Energy-Based Models (EBMs) provide an alternative by assigning an energy score to each event, which measures how well the event agrees with the learned data distribution. In this framework, a NN maps the input features into a latent space, where the energy is defined with respect to class-specific embeddings. This allows the model to identify noisy, anomalous or shifted events while maintaining good classification performance. We first study the method using controlled synthetic examples and then test it in a realistic collider analysis based on the $h\to\tau^+\tau^-$ process at the Large Hadron Collider (LHC), where $h$ is the discovered Standard Model (SM) Higgs boson. We show that the associated energy score can help identify events affected by mismodelling and/or systematic variations and can thus be used to define uncertainty-aware event selections. Although we use this specific Higgs process as a testbed, the method is general and can be applied to other collider classification problems where robustness against systematic and distributional uncertainties is important.
\normalsize{}
\end{abstract}
\newpage
\noindent\rule{\textwidth}{1pt}
\tableofcontents
\noindent\rule{\textwidth}{0.2pt}
\maketitle \flushbottom
\vspace{4mm}
\section{Introduction}

Deep Neural Networks (NN) have become an essential component of modern data analysis in High Energy Physics (HEP), with successful applications to particle identification, jet tagging, event classification, reconstruction and searches for physics Beyond the SM (BSM) \cite{Albertsson:2018maf,Guest:2018yhq}. Most supervised classifiers are trained in a discriminative manner, learning decision boundaries between pre-defined classes. Their performance, however, generally assumes that the training and inference samples are sufficiently representative of the same underlying distribution.

Jet tagging provides one of the clearest examples of the impact of Machine Learning (ML) at the LHC. The field has progressed from manually engineered substructure observables to increasingly low level representations of jets, including jet images \cite{Cogan:2014oua,Almeida:2015jua,deOliveira:2015xxd,Hammad:2022lzo,Esmail:2023axd}, particle clouds \cite{Komiske:2018cqr,Qu:2019gqs,Hammad:2024cae},  graph-based methods \cite{Verma:2020gnq,Abdughani:2018wrw,Moreno:2019bmu,Dreyer:2020brq,Esmail:2024jdg,Hammad:2025ewr,Fuks:2025qgh}, Lorentz equivariant networks \cite{Spinner:2024hjm,Favaro:2025pgz,Brehmer:2024yqw,Bogatskiy:2022czk}, and transformer-based architectures \cite{Qu:2022mxj,Esmail:2025kii,Hammad:2023sbd,Legge:2025cnm,Finke:2023veq,Hammad:2024qme}. These methods have achieved excellent performance in quark/gluon, heavy flavour, top, Electro-Weak (EW) boson, and Higgs tagging, while also increasing the importance of understanding classifier behaviour away from the training distribution.

A well known limitation of conventional discriminative classifiers arises when the training and inference distributions differ. NNs can assign high confidence to inputs that lie far from the training distribution or even outside its support \cite{nguyen2015deep}. In addition, modern neural classifiers can be poorly calibrated, such that the reported confidence does not always correspond to the true probability of a correct prediction \cite{guo2017calibration}. Softmax normalisation enforces relative probabilities among the known classes, but does not directly determine whether an input is compatible with the learned data distribution.

This limitation is particularly relevant at the LHC, where ML models are predominantly trained using Monte Carlo (MC) simulations. Although modern event generators and detector simulators for HEP provide sophisticated descriptions of collider events, differences between simulation and data remain unavoidable. These discrepancies may arise from the modelling of the hard process, parton shower, hadronisation, and/or underlying event, as well as from detector calibration, jet energy scale and resolution, reconstruction efficiencies,  and pileup. Highly expressive classifiers may exploit subtle features of the simulated distributions and can therefore become sensitive to such mismodelling.

Several strategies have been developed in collider physics to reduce the dependence of ML  classifiers on nuisance parameters and imperfect simulations. Adversarial training can be used to construct classifiers whose outputs are approximately insensitive to nuisance parameters, thereby improving robustness against systematic variations \cite{Louppe:2016ylz}. Related studies have explicitly demonstrated that deep classifiers can be sensitive to modelling choices in MC event generators, for example, through uncertainties associated with the parton shower \cite{Barnard:2016qma}. Weakly supervised approaches provide an alternative approach by reducing the reliance on event-by-event labels obtained from simulation, using instead mixed samples or data driven class information \cite{Dery:2017fap,Metodiev:2017vrx}. These ideas have also motivated model-independent  anomaly detection strategies for collider searches, in which the goal is to retain sensitivity to signals that are not represented by a specific simulated signal hypothesis \cite{Collins:2018epr,Heimel:2018mkt,Cerri:2018anq,Nachman:2020lpy}. Such methods have subsequently been applied directly to LHC data, including the ATLAS ones used in weakly supervised searches for resonant BSM physics \cite{ATLAS:2020iwa}.

These considerations motivate learning paradigms that characterise not only class boundaries but also the compatibility of an input with the learned data distribution. EBMs provide a natural framework for this purpose \cite{dawid2024introduction,Song:2021EBM}. Instead of relying only on normalised class probabilities, EBMs assign a (scalar) energy to a given configuration, with lower energies corresponding to configurations that are more compatible with the learned distribution and a higher energy otherwise, the latter thus indicating less typical configurations. Energy-based formulations have been widely studied in generative modelling, representation learning, and out-of-distribution (OOD)  detection \cite{du2019implicit,grathwohl2019your,liu2020energy}

This property is particularly attractive in HEP, where events encountered during inference stage may differ from the simulated training samples because of detector effects, theoretical modelling uncertainties, or previously unseen physical processes. Here, the energy score provides an additional measure of compatibility with the learned representation and can therefore be used to study classification reliability under distributional shifts and systematic variations.

In this work, we use the energy score not only as an OOD indicator, but also as a continuously adjustable selection variable. Rather than choosing an arbitrary threshold on a conventional classifier output, the energy threshold is calibrated directly against the reference-noise distribution. Each operating point therefore corresponds to a controlled reference-noise acceptance rate. Loosening the threshold increases the accepted signal yield at the cost of a larger noise contribution, while tightening it provides a purer but more restrictive selection.

We emphasise that this calibrated noise-acceptance rate is not identical to the purity of the selected sample, which also depends on the signal efficiency and on the relative signal and background rates. Nevertheless, such a energy-based calibration provides a transparent connection between the selection threshold and the tolerated noise contribution. This allows the analysis to be tuned to a desired operating point, e.g., by maximising the signal yield subject to a bound on noise acceptance or by optimising the expected statistical significance in the presence of systematic uncertainties.

The framework developed in this work therefore combines discriminative information with a distribution-aware energy score to provide an uncertainty-aware strategy for event selection. We investigate whether the learned energy landscape remains informative under distributional shifts and/or systematic variations as well as whether a threshold calibrated against the noise distribution can provide a more controlled operating criterion than conventional discriminative scores.

The plan of the paper is as follows. In the next section, we describe our EBM framework and illustrate it in the case of a toy model. After which, we move on to describe its application to LHC events. Ensuing results are then presented, before summarising and concluding.
\section{EBM Description}
\label{sec:model}

EBMs constitute a broad class of probabilistic learning frameworks in which the behaviour of a system is characterised through an energy function rather than directly through normalised probabilities. Given an input $\vx$ and an associated label $y$, an EBM assigns a scalar energy value
\begin{equation}
E_\theta(\vx,y),
\end{equation}
where $\theta$ denotes the trainable parameters of the model. The central principle of energy-based learning is that configurations compatible with the underlying data distribution should be assigned lower energies, whereas unlikely or incompatible configurations should correspond to higher energies. In this sense, the energy function can be viewed as a measure of compatibility between an observation and a candidate label.
The probabilistic interpretation of EBMs follows naturally from statistical mechanics. The energy function induces a Gibbs distribution
\begin{equation}
p_\theta(\vx,y) =
\frac{\exp[-\beta E_\theta(\vx,y)]}
{Z_\theta},
\label{eq:gibbs_joint}
\end{equation}
where
\begin{equation}
Z_\theta =
\int dy'
\exp[-\beta E_\theta(\vx,y')]
\label{eq:partition}
\end{equation}
is the partition function that ensures proper normalisation of the probability distribution.
For classification tasks, the conditional probability of a class label can be obtained from the energy function through
\begin{equation}
p_\theta(y|\vx) =
\frac{\exp[-\beta E_\theta(\vx,y)]}
{\sum\limits_{y'}
\exp[-\beta E_\theta(\vx,y')]}.
\label{eq:conditional}
\end{equation}
Unlike conventional discriminative classifiers, which directly learn the conditional probability distribution $P(y|\vx)$, EBMs first construct an energy landscape over the joint input--label space and subsequently derive probabilities from this landscape. The predicted label is then determined according to the minimum energy principle,
\begin{equation}
\hat y =
\arg\min_y E_\theta(\vx,y).
\label{eq:min_energy}
\end{equation}

A distinctive feature of EBMs is that the energy itself provides information beyond the final classification decision. While conventional softmax classifiers only encode relative preferences among the available classes, the magnitude of the energy can be interpreted as a measure of compatibility between an event and the learned data distribution. Consequently, events that are poorly described by the training distribution tend to occupy regions of higher energy. This property makes EBMs naturally suited for uncertainty quantification, anomaly detection, and OOD identification.

From a geometric perspective, the training procedure aims to shape the energy landscape such that regions populated by data correspond to low energy basins separated by higher-energy barriers. During inference, classification is achieved by locating the class associated with the lowest energy configuration while the absolute energy value provides a quantitative measure of how well the event is represented by the learned manifold. These characteristics distinguish EBMs from purely discriminative approaches and have motivated their increasing application to problems where robustness, uncertainty estimation, and anomaly detection play a central role.

\subsection{Proposed Structure}
\label{sec:structure}
In this work, we formulate the classification problem within the framework of EBMs, where predictions are derived from an energy function that quantifies the compatibility between an input event and a given class label. Let $\vx \in \mathbb{R}^{d}$ denote the input feature vector and $y \in [0,1]$ the corresponding binary label. Rather than directly learning class probabilities, the model assigns an energy value
\begin{equation}
\Energy(\vx,y) =
\left|f_{\theta}(\vx)-\ve_y\right|^2 ,
\label{eq:energy}
\end{equation}
where $f_{\theta}:\mathbb{R}^{d}\rightarrow\mathbb{R}^{m}$ is a NN that maps the input into a latent representation space and $\ve_y\in\mathbb{R}^{m}$ is a learnable embedding associated with class $y$. The energy is defined as the squared Euclidean distance between the latent representation and the corresponding class embedding. Consequently, events whose latent representations lie close to a class prototype are assigned low energy while incompatible events receive larger energy values.
The feature extractor $f_{\theta}$ is implemented as a fully-connected NN,
\begin{equation}
f_{\theta}(\vx)
=
W_{L},\sigma\Bigl(
W_{L-1}\cdots
\sigma(W_{1}\vx+b_{1})
\cdots+b_{L-1}
\Bigr)+b_{L},
\label{eq:extractor}
\end{equation}
where $W_i$ and $b_i$ denote the trainable weights and biases of the NN, respectively. Each hidden layer is followed by Layer Normalisation, a LeakyReLU activation function, and Dropout regularisation.
This architecture was found to provide sufficient representational power while maintaining computational efficiency and training stability.
The energy function naturally induces a probabilistic interpretation through the Gibbs distribution,
\begin{equation}
p_{\theta}(y|\vx)
=
\frac{
\exp\left[-\beta\Energy(\vx,y)\right]
}
{
\sum\limits_{y'}
\exp\left[-\beta\Energy(\vx,y')\right]
},
\label{eq:gibbs}
\end{equation}
where $\beta$ denotes a learnable inverse temperature parameter controlling the sharpness of the resulting probability distribution. Since the label space is finite, the partition function can be evaluated exactly, allowing the model to be trained using standard optimisation techniques. In contrast to conventional discriminative classifiers, however, the energy function remains explicitly accessible and provides additional information beyond class probabilities.

Classification is performed according to the minimum energy principle,
\begin{equation}
\hat{y}(\vx)
=
\arg\min_{y\in\mathcal{Y}}
\Energy(\vx,y),
\label{eq:predict}
\end{equation}
which assigns each event to the class with the lowest energy. Beyond classification, the energy landscape enables a direct assessment of how well an event is described by the learned data distribution. To quantify this compatibility, we define the energy score
\begin{equation}
S(\vx)
=
\min_{y\in\mathcal{Y}}
\Energy(\vx,y),
\label{eq:noise_score}
\end{equation}
which corresponds to the minimum energy attained by the event over all possible classes. Events associated with large values of $S(\vx)$ are poorly represented by the learned latent manifold and can therefore be interpreted as noisy samples. In practice, a threshold $\tau$ is determined using the training data and events satisfying $S(\vx)>\tau$ are flagged as potential anomalies.
The threshold $\tau$ is calibrated using a reference sample representative of the
training distribution. A common choice is to define $\tau$ as a high percentile
of the energy-score distribution,
\begin{equation}
P\bigl(S(\vx)\le \tau\bigr)=1-\alpha,
\end{equation}
where $\alpha$ denotes the desired false positive rate. For example, choosing $\alpha=0.01$ implies that approximately $1\%$ of events drawn from the reference distribution are classified as anomalous. Events satisfying $S(\vx)>\tau$ are therefore identified as potential noise samples.

The proposed EBM therefore provides a unified framework for event classification and uncertainty-aware anomaly identification. While the relative class energies determine the predicted label, the absolute minimum energy provides an independent measure of how compatible an event is with the learned data distribution. This allows the same model to perform classification while simultaneously identifying events that lie outside the training manifold.
\subsection{Training Objective}
\label{sec:training}

The parameters of the energy function are optimised by minimising the negative
log-likelihood (NLL) associated with the Gibbs distribution. For a training sample consisting of $N$ labeled events,
the corresponding objective function is given by
\begin{equation}
  \mathcal{L}(\theta)
  =
  \frac{1}{N}
  \sum_{i=1}^{N}
  \left[
    \Energy(\vx_i,y_i)
    +
    \log
    \sum_{y'}
    \exp\!\left(-\Energy(\vx_i,y')\right)
  \right]\,,
  \label{eq:loss}
\end{equation}
with $y_i$ denotes the true class label associated with the $i$-th training event, while $y'$ is a dummy index that runs over the two possible class labels, $y'=0,1$.
The first term favours low energy configurations for the correct class assignment, while the logarithmic partition term suppresses competing class hypotheses by increasing their relative energies. As a result, the optimisation procedure shapes the energy landscape such that events belonging to the training distribution occupy low energy regions, whereas incompatible configurations are assigned larger energy values, as intended. This separation is essential for the identification of noisy events, as samples that are poorly described by the learned distribution naturally acquire elevated energies.

To quantify the compatibility of an event with the learned data manifold, we employ the energy score defined in equation~\eqref{eq:noise_score}. Following training, a threshold $\tau$ is determined from an independent validation sample representative of the training distribution. Specifically, $\tau$ is chosen as the $p$-th percentile of the corresponding energy score distribution,
\begin{equation}
  \tau 
  =
  \operatorname{Quantile}_{p}
  \!\left(
    \left\{
    S(\vx_i)
    \;:\;
    (\vx_i,y_i)\in\mathcal{D}_{\rm val}
    \right\}
  \right),
  \label{eq:threshold}
\end{equation}
where $\mathcal{D}_{\rm val}$ denotes the validation dataset. In this work, we adopt $p=99$, corresponding to a false positive rate of approximately $1\%$ on the validation sample. This calibration procedure introduces no additional trainable parameters and requires only a single evaluation of the energy score on the validation dataset.

During inference, each event is first evaluated according to its minimum energy score. Events satisfying
\begin{equation}
  S(\vx) > \tau
\end{equation}
are flagged as noisy event candidates, while the remaining events are assigned a class label according to the minimum energy criterion,
\begin{equation}
  \hat{y}(\vx)
  =
  \arg\min_{y\in\mathcal{Y}}
  \Energy(\vx,y).
  \label{eq:decision}
\end{equation}

In this case noisy events typically acquire large energy values irrespective of the relative separation between classes, allowing them to be identified through the global structure of the learned energy landscape.

\subsection{Energy Score and Uncertainty-Aware Selection}
\label{sec:uncertainty_selection}
 
While Section~\ref{sec:training} introduced the threshold $\tau$ as a fixed quantile for identifying noisy events, the same threshold can also be used to define the event-selection criterion. For a given percentile $p$, the corresponding threshold $\tau_p$ is determined from equation~\eqref{eq:threshold} and all events satisfying $S(\mathbf{x}) \leq \tau_p$ are accepted.
By construction, the acceptance region contains a fraction $p/100$ of the reference (training-like) distribution. The remaining fraction,
\begin{equation}
\alpha(\tau_p)=1-\frac{p}{100},
\end{equation}
represents the probability that a reference-like event lies above the threshold. When $\tau_p$ is applied to an independent target sample, $\alpha(\tau_p)$ provides a calibrated estimate of the expected contamination from noisy events.
Scanning the percentile $p$ therefore generates a family of selection thresholds, each corresponding to a different trade-off between signal efficiency, $\varepsilon_{\rm sig}$, and the expected noise contamination, $\alpha$. Unlike a conventional classifier, where the operating point is chosen by applying an arbitrary cut on the classifier score, each energy threshold has a direct statistical interpretation through its associated contamination rate.

\begin{figure}[!h]
  \centering
  \includegraphics[width=0.65\linewidth]{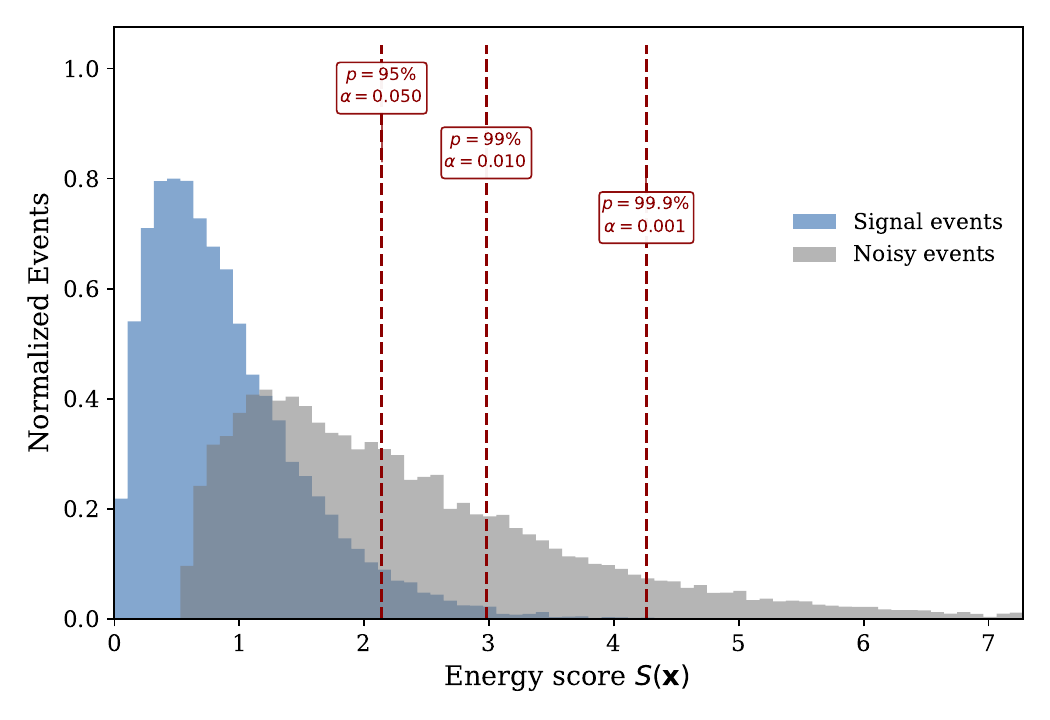}
  \caption{
    {Energy score distributions with tunable operating points.}
    Signal (blue) and noise (grey) energy score distributions, with
    thresholds $\tau_p$ marked for several candidate percentiles $p$ and
    their corresponding calibrated uncertainty $\alpha = 1-p/100$. Loosening
    $\tau_p$  accepts more of the noise tail but also more of
    the signal distribution beyond the conventional cut.
  }
  \label{fig:energy_operating_points}
\end{figure}

This provides two important advantages for collider analyses. First, the selection threshold can be adjusted after training without retraining the model, allowing the analysis to maximise the retained signal while satisfying a chosen upper limit on the accepted noise contamination. Second, because $\alpha(\tau)$ is determined directly from the energy-score distribution, it does not depend on detailed knowledge of the noise composition. Consequently, the method remains applicable even when the noise originates from detector effects, imperfect simulations, and/or previously unseen event topologies.

We emphasise that this use of the energy score differs from the conventional application of EBMs to OOD rejection discussed above. In the standard approach, events that are inconsistent with the training distribution are simply identified and removed. Here, the energy threshold $\tau$ is used as a tunable analysis parameter. By scanning over different values of $\tau$, one can determine the operating point that retains the largest amount of signal while satisfying a chosen level of uncertainty in the sample purity. Figure~\ref{fig:energy_operating_points} illustrates this procedure by showing the energy score distributions for the signal and noise samples together with several candidate thresholds $\tau_p$.

\subsection{Illustration with a Toy Model}
\label{sec:experiments}
For further illustration, we construct a synthetic binary classification problem in $d=20$ dimensions using $10{,}000$ clean samples in total, equally divided between the two classes. Ten dimensions are chosen to carry class-discriminating information, while the remaining ten dimensions contain only weak, non-informative fluctuations.

In the informative subspace, the two classes are sampled from isotropic Gaussian distributions with a per-coordinate standard deviation of $\sigma=0.8$. Their centres are located at $0$ and $3.1$, respectively, along each informative coordinate, corresponding to a centre-to-centre separation of
$3.1\sqrt{10}\simeq 9.8.$
The remaining ten dimensions carry no class information and are sampled from Gaussian distributions with $\sigma=0.1$.
To emulate anomalous or noisy events, we additionally generate $2{,}500$ samples, corresponding to $25\%$ of the clean sample size. These events are drawn from a broader Gaussian distribution with $\sigma=4$, whose mean is displaced by $\pm4$ along each coordinate, with the signs chosen randomly. This construction populates regions of feature space that are typically away from the two clean class centres. The noise samples are assigned random binary labels.
The full dataset therefore contains $12{,}500$ events. After shuffling, $80\%$ of the samples are assigned to the training pool and the remaining $20\%$ to the test set. The test set contains $2{,}500$ events, of which $496$ are noise samples. The corresponding training pool therefore contains $2{,}004$ noise events and $7{,}996$ clean events. Only the clean component of the training pool is used for model optimisation. Of these clean events, $7{,}197$ are used for training and $799$ are reserved for validation and threshold calibration. The model therefore does not encounter noise samples during training.
Finally, all input features are standardised using the mean and standard deviation computed from the clean training sample only.

\begin{table}[!h]
  \centering
  \renewcommand{\arraystretch}{1.25}
  \begin{tabular}{lc}
    \toprule
    \textbf{Metric} & \textbf{Value} \\
    \midrule
    Classification accuracy (clean samples)      & $98.4\%$ \\
    Noise detection accuracy                      & $97.1\%$ \\
    Noise true positive rate (recall)             & $96.3\%$ \\
    Noise true negative rate (specificity)        & $97.3\%$ \\
    Noise detection AUC                           & $0.993$  \\
    \bottomrule
  \end{tabular}
   \caption{Quantitative results on the held-out test set.}
   \label{tab:results}
\end{table}

\begin{figure}[!ht]
  \centering
  \includegraphics[width=\linewidth]{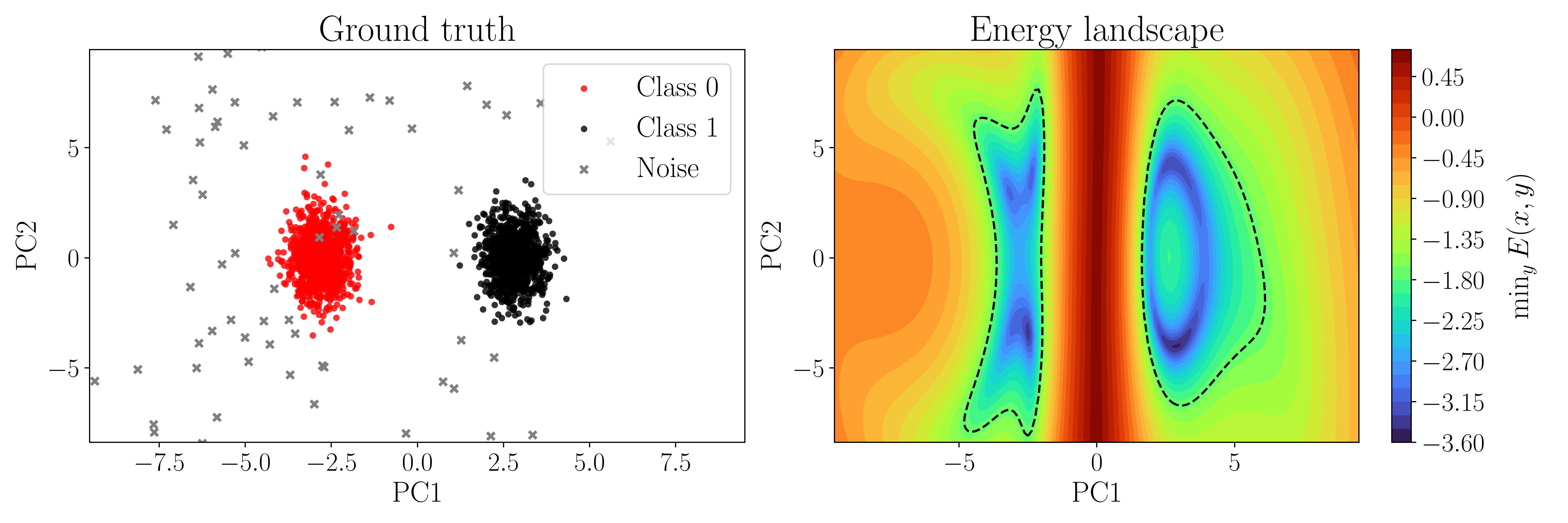} \\
  \includegraphics[width=0.85\linewidth]{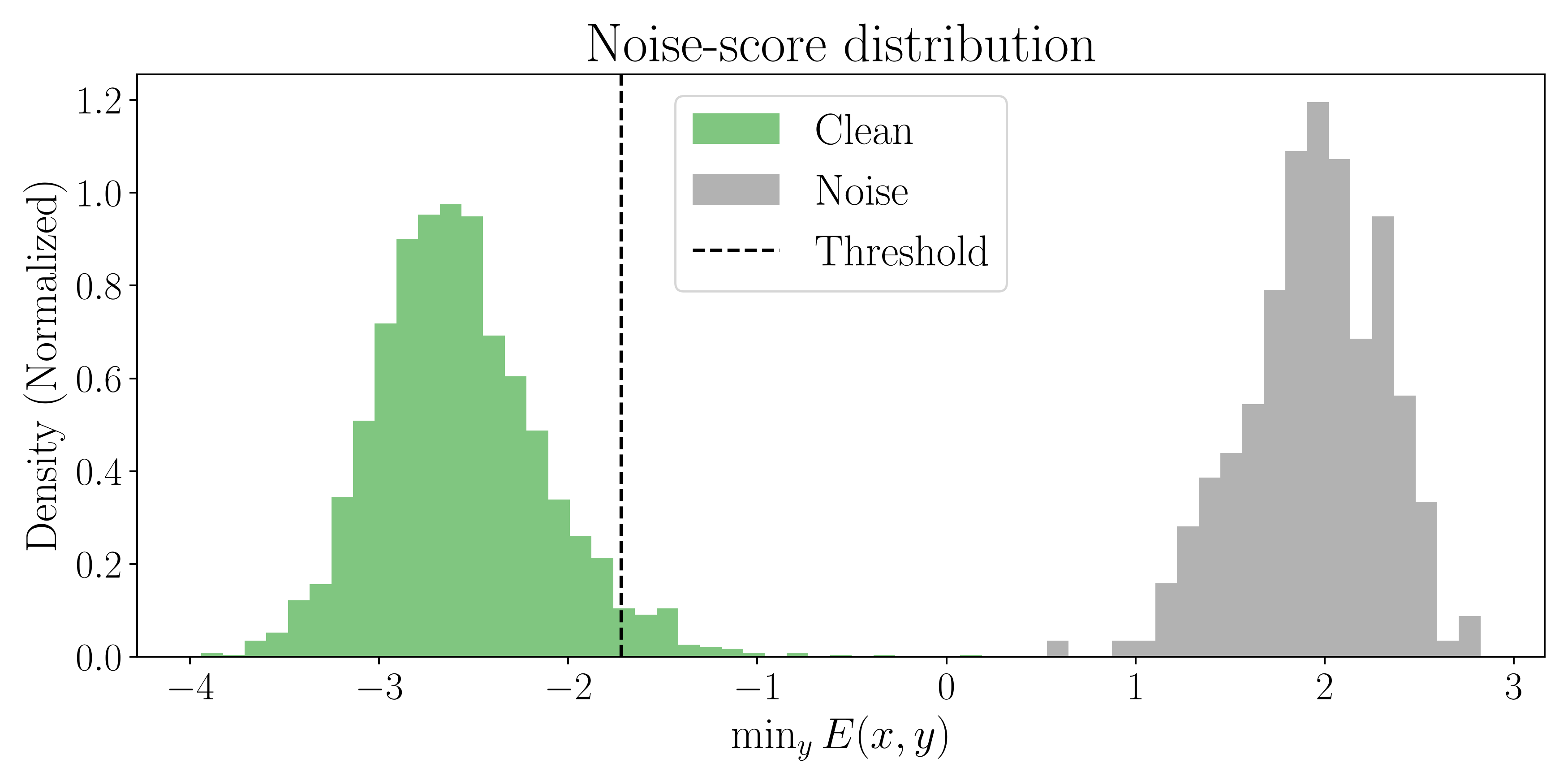}
  \caption{
    {Top-Left:}~Test samples in the plane of the first two principal components of the
    clean test data: class~0 (red), class~1 (orange) and noise (gray crosses).
    {Top-Right:}~Learned energy landscape, shown as  $S(\vx) = \min_y E(\vx, y)$. The dashed black contour is the noise boundary
    $S(\vx) = \tau$; samples outside it are flagged as noise.
    {Bottom:}~Distribution of  $S(\vx)$ for clean (green) and noise (gray) test samples: here, the dashed line marks the threshold $\tau$.
     }
  \label{fig:toy}
\end{figure}
 
Training converges quickly and smoothly, the loss falls from $1.38$ to $0.03$ within the first three epochs and reaches $5\times10^{-4}$ at the end, with the validation loss following the training loss closely. The absence of oscillations or divergence, common failure modes of EBM training with
Markov Chain MC (MCMC)-based negative sampling, follows from the tractable partition function provided by
the finite label set.  
Table~\ref{tab:results} lists the performance on the test set. With a centre-to-centre distance of $12\,\sigma$ the two classes are well separated, and all $2{,}004$ clean test samples are classified correctly: the energy difference $E(\vx,0) - E(\vx,1)$ equals $-20.9 \pm 0.7$ for class~0 and $+20.9 \pm 0.5$ for
class~1.
 
Figure~\ref{fig:toy} shows the test data, the learned energy landscape and the distribution of the noise score. For visualisation, the data are projected onto the first two principal components of the clean test samples and the energy is evaluated on a grid in this plane that is mapped back to the 20-dimensional input space.

Two observations stand out in the energy landscape, top frames of Figure~\ref{fig:toy}. First, the noise boundary forms two compact closed regions, one around each class cluster, so samples far from both classes are rejected {without any supervised signal from noise samples}. Second, the two low energy regions are separated by a high energy ridge along the class boundary, where $E(\vx,0) \approx E(\vx,1)$. A sample therefore receives a low energy only if it lies close to the training data of one class. The shape of the noise boundary follows the geometry of the learned energy, rather than an ad-hoc cut on a classifier output.

The bottom panel of Figure~\ref{fig:toy} shows that the clean and noise score distributions are fully disjoint. The median noise score exceeds the median clean score by more than four orders of magnitude and the lowest scoring noise sample lies a factor of about three above the highest scoring clean sample.
Consequently, the ranking given by $S(\vx)$ is perfect, with an Area Under the Curve  (AUC)  equal to 1, and every remaining error coming from the choice of $\tau$. The $81$ rejected clean samples  are consistent with calibrating $\tau$ at the $95^{\text{th}}$ percentile, which by construction gives up about $5\%$ of clean data for a conservative noise boundary, they lie in the outer tails of the two clusters and none of them would have been misclassified.
Any threshold in the gap between the two distributions ($1.2 < \tau < 3.5$, compared with the calibrated $\tau = 0.019$) would give zero errors on this test set, so the result is robust to the calibration of the threshold.

\section{Application to Systematic Uncertainty in Collider Events}
In this section we apply the idea of EBMs to collider events that have been affected by systematic effects from the detector. Typically, simulations of signal and background processes are performed assuming nominal performance of the detector and used to train NNs with high discriminative power. However, when applied to real events,  systematic effects shift the event information used for classification resulting in performance that is highly dependent on the degree of this systematic effects. Our goal and procedure follows the approach adopted in Ref.~\cite{Ghosh:2021roe}, where a set of different network architectures are tested against collider events  with a simulated nuisance parameter.
In our case, we explore the possibility of tuning the energy function as well as the probability distribution with the objective of improving the signal strength estimation.

For the evaluation of the NN we use our EBM to create templates of signal and background and calculate a binned likelihood $\mathcal{L}(\mu, z, | \{\vx_i\})$.
This binned likelihood contains a Poisson term for the counts of the events in the bins and a term for the constraint on the nuisance parameter $z$. The corresponding NLL is given by 
\begin{align}
    \label{eq:nll}
    -\log \mathcal{L}(\mu, z, | \{\vx_i\}) = {}&
    - \sum_{j=1}^{n_\text{bins}} \left[N_j \log(\mu s_j + b_j) - \mu s_j - b_j - \log(\Gamma(N_j))\right]\nonumber\\
    &- \log\left(\frac{1}{\sigma_z \sqrt{2\pi}}\right) - \left(\frac{z - z0}{\sqrt{2} \sigma_z}\right)^2 \,,
\end{align}
where $j$ is the bin index, $s_j$ and $b_j$ are expected numbers events for signal and background, respectively, $N_j$ is the number of events in data,
all of these in their corresponding bin. The function $\Gamma$ is the well-known extension of the factorial function to the complex numbers.
In equation~\eqref{eq:nll}, $\mu$ is the signal strength that we will try to fit by minimizing the NLL.

The EBM we employ for the classification of collider events follows equation~\eqref{eq:energy}.  The object to be trained is the feature extractor $f_\theta$ that will take care of mapping the inputs $\vx_i$ to an $n$-dimensional embedding.  During training, the loss function uses the energy of equation~\eqref{eq:energy} such that in this embedding background events are moved closer to $\ve_0$ and away from $\ve_1$, and vice versa for signal events. Training is done using the log-loss function given by:
\begin{equation}
    \label{eq:logloss}
    \mathcal{L}_\text{log} (\theta) = \frac{1}{N} \sum_{i=1}^{N} \log \bigl[
        1 + \exp \bigl( E_\theta (\vx_i, y_i) - E_\theta (\vx_i, \bar{y}_i) \bigr)
    \bigr]
\end{equation}
where $\bar{y}_i$ represents the wrong class for event $\vx_i$.
It becomes apparent that the loss function given above is a good choice for this example
if we write it in terms of probabilities.
Using equation~\eqref{eq:gibbs} with $\beta = 1$,
it is easy to show that
$\mathcal{L}_\text{log} (\theta)$
is equivalent to
\begin{equation}
    \label{eq:loglossprob}
    \mathcal{L}_\text{log} (\theta) = -\frac{1}{N} \sum_{i=1}^{N} \log \bigl[
        p_\theta (y_i | \vx_i)
    \bigr] \,.
\end{equation}

Typically, after completing the training described above, the trained NN can be used to perform the numerical analysis. However,
in the case of EBM we obtain energies that provide additional information, useful to perform a number of preliminary adjustments before carrying out the final analysis. In this section we explore different strategies for calibrating the inverse temperature, $\beta$, which determines the signal and background scores
from energy values through equation~\eqref{eq:gibbs}. The first thing to note in this section, is that tuning the $\beta$ parameter will only reshape the distribution of the scores moving the peaks towards or away from the boundaries~\cite{Mattei2026}. In other words, it will not correct misclassified points
but will move their scores. 
By training the network first, and tuning the inverse temperature later, we are practically turning the model into an adjustable one-dimensional classifier,
easy and fast to train as well as very difficult to overfit. In the usual case, one would be concerned with making sure that training results in a model with high confidence when predictions are correct while keeping confidence low when predictions are wrong. In our case (two class discrimination) cross-entropy ensures that confidences follow that rule. Interestingly, cross-entropy can be formulated using NLL and the corresponding loss function of $\beta$ would be
\begin{equation}
    \label{eq:betace}
    \mathcal{L}_\mathrm{ce}(\beta) = - \frac{1}{N} \sum_{i=1}^N \log[p_\beta (y_i|\vx_i)],
\end{equation}
which seems identical to equation~\eqref{eq:loglossprob}, except for the dependence on $\beta$ instead of the network parameters $\theta$. The optimised value is then obtained such that $\beta_\mathrm{ce}  = \mathrm{argmin}_{\beta > 0} \mathcal{L}_\mathrm{ce}(\beta)$. 

An important point we want to make here is that we can employ the energies to perform the optimisation of $\beta$ in any way we may want to. Therefore, we can perform this optimisation with the objective of improving the analysis performed in this section. The simplest idea would be to tune $\beta$ such that template shapes change as little as possible between values of the nuisance parameter, and that their shapes resemble as much as possible the most accurate template.
Obviously, here we are still talking about templates of a training dataset. Considering that we want to improve the estimation of $\mu$ and that this factor comes in front of signal bin heights, one loss function that could do this optimisation is
\begin{equation}
    \label{eq:betasn}
    \mathcal{L}_\mathrm{sn}(\beta) = \sum_{j=1}^{n_z} \sum_{i=1}^{n_\mathrm{bins}} \left| s_{i,z_j}(\beta) - s_{i,z_\mathrm{ref}}(\beta)\right|^\gamma ,
\end{equation}
where $s_{i,z}(\beta)$ are bin heights obtained from probability scores calculated using energies from the NN, $z_\mathrm{ref}$ is some reference $z$ value for which the network has the best accuracy and confidence, and the sums run over all the $n_\mathrm{bins}$ bins and the $n_z$ nuisance parameter $z$ values.
The exponent $\gamma$ can be used to adjust the steep of this loss function. The optimised value corresponds to the same condition $\beta_\mathrm{sn}  = \mathrm{argmin}_{\beta > 0} \mathcal{L}_\mathrm{sn}(\beta)$ The use of bin heights in this loss function make it non-differentiable. This is not a problem since performing optimisation with one variable is easy and fast with any scalar optimizer that does not require gradient values. Again, this is based on the idea that, while $\beta$ optimisation does not improve accuracy, it changes the shapes of the probability distribution and we are attempting to obtain shapes that improve our final calculation.

\begin{figure}[!h]
    \centering
    \includegraphics[width=\linewidth]{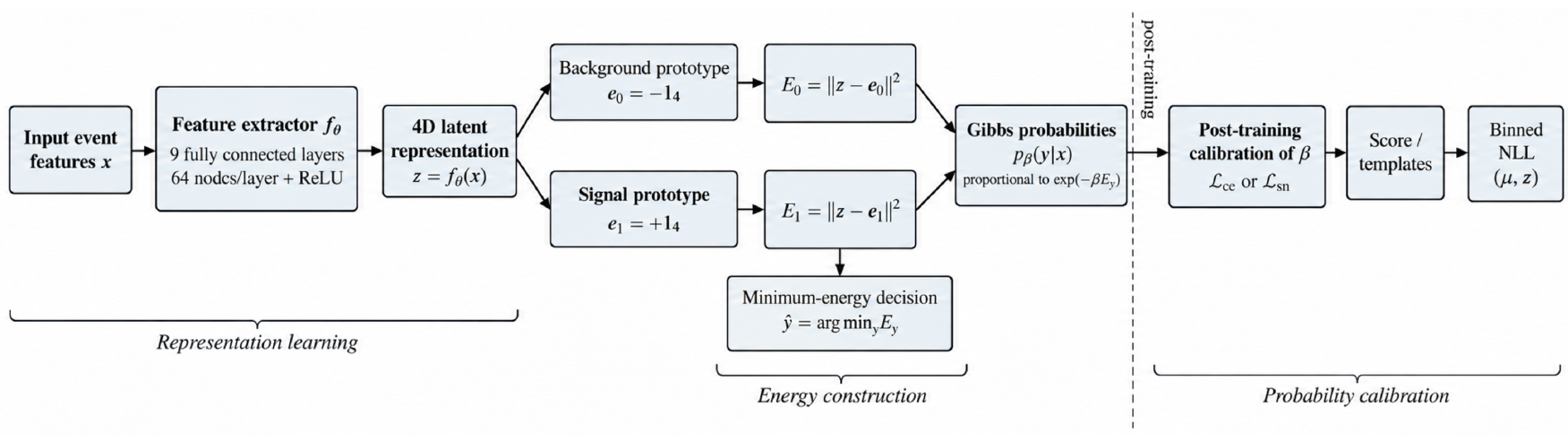}
    \caption{Schematic representation of the EBM used in the collider analysis. The input features are mapped to a latent representation, from which class energies, probabilities,  and the calibrated output are obtained.}
    \label{fig:network}
\end{figure}

The complete workflow is summarised in figure ~\ref{fig:network}. The input event features are first mapped by the feature extractor $f_\theta$ into a low-dimensional latent representation. The latent vector is compared with the signal and background class embeddings while the corresponding squared distances define the class energies. These energies determine the minimum energy classification and are converted into signal and background probabilities through the Gibbs distribution.

A useful feature of this construction is the separation between representation learning and probability calibration. The parameters of $f_\theta$ are optimised first using the energy-based classification loss while the inverse temperature $\beta$ is calibrated only after the network has been trained. Consequently, the learned latent representation and energy ordering remain unchanged, whereas the probability score distributions can be reshaped according to the final objective of the analysis. In the present application, we consider both the conventional cross-entropy calibration and the signal template calibration introduced above.

\section{Results on the HiggsML Dataset}
\label{sec:higgsmlresults}
With the intention of directly comparing against Ref.~\cite{Ghosh:2021roe}, we closely follow their procedure on the same dataset. We evaluate the performance of the process detailed in the previous section using the HiggsML dataset shared by ATLAS~\cite{pmlr-v42-cowa14}, a simulated dataset for the decay of the Higgs to a pair of $\tau$ leptons. We consider different values for a nuisance parameter that characterises systematic effects on the hadronically decaying $\tau$ energy scale and test the robustness of EBMs to the classification of events under these effects.

The dataset was generated by the ATLAS collaboration using their official full detector simulator. As hinted above, the signal in this dataset consist of events
in which a Higgs boson decays to a pair of $\tau$.
This Higgs is produced
via gluon-gluon fusion,
vector boson fusion,
or together witha a vector boson.
The background correspond to other channels that may produce an hadronic $\tau$.
For the challenge, the backgrounds include the decay of the $Z$ boson,
events with a pair of top quarks,
and events decaying $W$ bosons~\cite{pmlr-v42-cowa14}.

As it was done in Ref.~\cite{Ghosh:2021roe}, we employ the dataset from the HiggsML Kaggle challenge by ATLAS, with a skewing  function~\cite{victor_estrade_2018_1887847} using different values for a nuisance parameter $z$. The use of a skewing functions follows the approach of previous studies~\cite{Estrade2019}. (For more details on the effects of this nuisance parameter on the features of the dataset see Sec.~V of Ref.~\cite{Ghosh:2021roe}.)
Following the recommendation of previous works, we consider a minimum threshold of $E_T > 22$~GeV after applying the skewing function to the events.
We split the data in 2/3 for training and 1/3 for testing. Furthermore, 2/3 of the training data will be used to train the encoder $f_\theta$,
while 1/3 of the training data will be used for training the inverse temperature $\beta$.

The feature extractor $f_\theta$ consists of nine fully connected hidden layers, each containing 64 nodes and using a Rectified Linear Unit (ReLU) activation function. The final layer is a linear layer with four outputs, mapping each input event $\vx$ to a four-dimensional representation. The background ($y = 0$) and signal ($y = 1$) classes are embedded in two vectors $\ve_y$ with $\ve_0 \neq \ve_1$. These two vectors can be decided during training or can be fixed before training. In our case, fixing them before training improved stability of the optimisation. We set all the elements of $\ve_0$ and $\ve_1$ to $-1$ and 1, respectively. Thus, during training the model must learn a latent space that for background events is closer to $\ve_0$ and away from $\ve_1$, and vice versa for signal events. Training is done using the log-loss function given in equation~\eqref{eq:logloss}.

\begin{figure}[!htbp]
  \centering
  \includegraphics[width=0.4\linewidth]{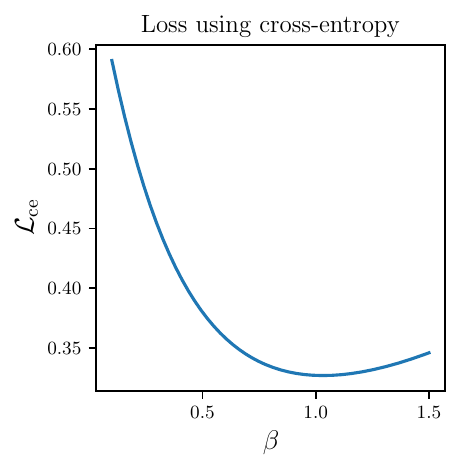}%
  \includegraphics[width=0.4\linewidth]{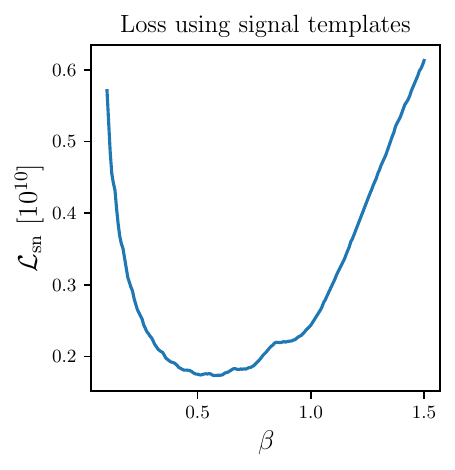}
  \caption{
      Dependence of the loss functions
      $\mathcal{L}_\mathrm{ce}$ of equation~\eqref{eq:betace}
      and $\mathcal{L}_\mathrm{sn}$ of equation~\eqref{eq:betasn} on the inverse temperature $\beta$.
  }
  \label{fig:lossbeta}
\end{figure}

\begin{figure}[!htbp]
  \centering
  \includegraphics[width=0.45\linewidth]{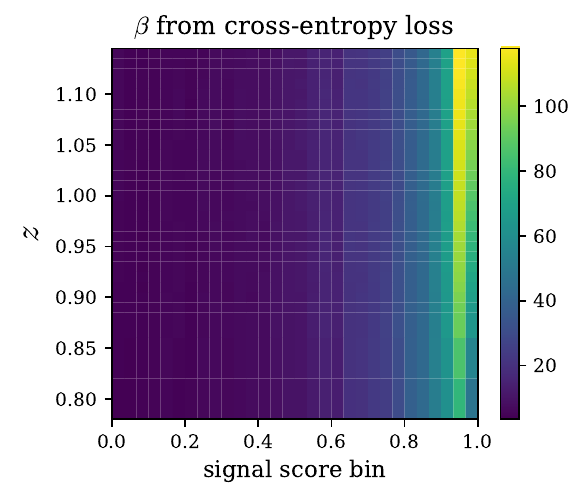}%
  \includegraphics[width=0.45\linewidth]{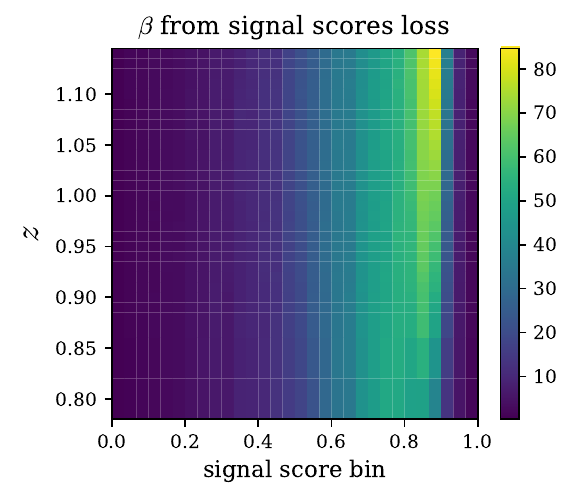}
  \caption{
      Binned signal scores for the probabilities obtained from equation~\eqref{eq:gibbs}
      using two different values of $\beta$.
      The panel on the left uses $\beta = 1.0322$ obtained from minimising equation~\eqref{eq:betace}
      while the panel on the right uses $\beta = 0.5927$ obtained from minimising equation~\eqref{eq:betace}.
      Color corresponds to the heights of binned scores.
  }
  \label{fig:sigscores}
\end{figure}

Training the model to minimise an energy of the form of equation~\eqref{eq:energy} benefits from having large batches such that distribution variations per class between batches is small. Therefore, we train using batches of $\sim 200,000$ events. For training we use the Root Mean Square Propagation (RMSProp) algorithm with a learning rate of $10^{-3}$, with a total of 50 training epochs. The setup of the EBM is summarised in Table~\ref{tab:higgsmlebm}. After training the encoder, we obtain the corresponding energies and proceed to optimising the inverse temperature $\beta$. By minimising the cross-entropy loss function
of equation~\eqref{eq:betace} we obtain a value $\beta_\mathrm{ce} = 1.0322$ resulting in an slight increase in confidence. However, with the loss function for signal templates of equation~\eqref{eq:betasn}, with $\gamma = 2$, we obtain $\beta_\mathrm{sn} = 0.5927$, indicating that this loss function benefits from reduced confidence. The dependence of these loss functions on $\beta$ is shown in figure~\ref{fig:lossbeta} and the binned signal scores obtained from the two different values of $\beta$ are shown in figure~\ref{fig:sigscores}. One interesting feature from figure~\ref{fig:lossbeta}  is that the value of $\mathcal{L}_\mathrm{sn}$, the right panel, seems to oscillate, but this is easy to explain considering that it uses binned signal scores. In the same figure, the left panel, we show that $\mathcal{L}_\mathrm{ce}$ has a simple behaviour with a minimum close to 1.

\begin{table}[!t]
  \centering
    \caption{
       Hyperparameter configuration of the EBM network.
    The notation $\mathbf{1}_4$ stands for a 4 dimensional vector
    with all entries equal to 1.
    }
  \label{tab:higgsmlebm}
  \renewcommand{\arraystretch}{1.25}
  \begin{tabular}{lc}
    \toprule
      Hidden layers (HL)  & 9 \\
      Nodes per HL  & 64 \\
      Activation function in HL & ReLU \\
      Embedding dimensions & 4 \\
      Energy function & equation~\eqref{eq:energy} \\
      $\mathbf{e}_0$ & $-\mathbf{1}_4$ \\
      $\mathbf{e}_1$ & $\mathbf{1}_4$ \\
      Epochs & 50 \\
      Batch size & $\sim$ 200,000 \\
      Optimizer & RMSProp \\
      Learning rate & $10^{-3}$ \\
      Loss function ($f_\theta$ parameters) & equation~\eqref{eq:logloss} \\
      Loss function ($\beta$) & Equations~\eqref{eq:betace} and~\eqref{eq:betasn} ($\gamma = 2$) \\
    \bottomrule
  \end{tabular}
\end{table}

To compare against the result in Ref.~\cite{Ghosh:2021roe},  we repeat their configuration for the ``Uncertainty-aware classifier'', a model that joins two  subnetworks, one in charge of the systematic effects above and at nominal nuisance parameter, and another one in charge of systematic effects below nominal.
We will refer to this model as \emph{Conditional MLP} to avoid confusion. The subnetworks have 10 hidden layers with 64 nodes, ReLU activation function,
and a one-dimensional output with sigmoid activation function. The training uses RMSProp for optimisation with binary cross-entropy loss function,
with batches of size 4096. Regarding differences between our EBM and Ref.~\cite{Ghosh:2021roe}, we consider repeating their calculation with large batches
as in our case, but this results in overfitting. Additionally, to obtain a result as close as possible to the best when training the Conditional MLP,
we validate using the test dataset and choose the model parameters that give the best accuracy. Obviously, in our case, we only use the test dataset to evaluate the final result and never for validation.

\begin{figure}[!htbp]
  \centering
  \includegraphics[width=0.33\linewidth]{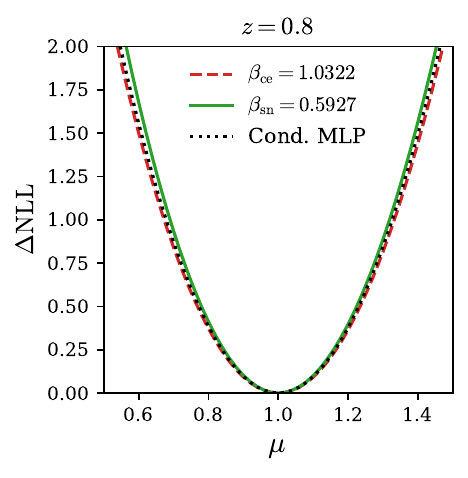}%
  \includegraphics[width=0.33\linewidth]{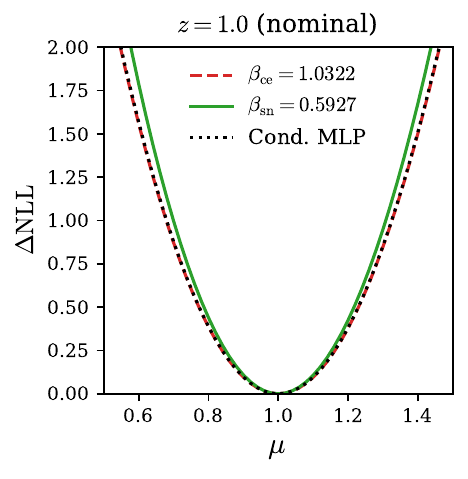}%
  \includegraphics[width=0.33\linewidth]{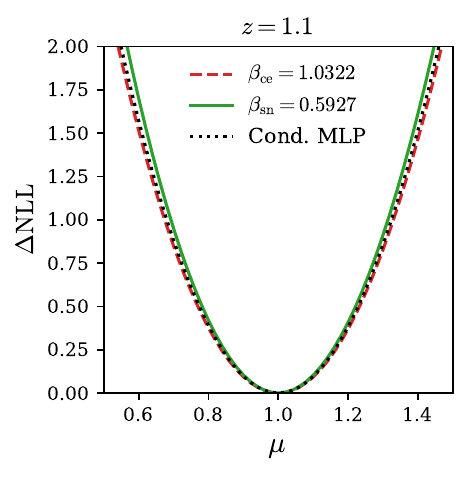}
  \caption{
      Different curves obtained for $\mu$
      when calibrating the inverse temperature
      using cross-entropy loss ($\beta_\mathrm{ce}$)
      and signal templates ($\beta_\mathrm{sn}$).
      A value of $\beta$ close to 1 (the default used during training)
      shows that training itself was already close to the optimal $\beta$
      for cross-entropy.
  }
  \label{fig:calibcompare}
\end{figure}

\begin{figure}[!h]
  \centering
  \includegraphics[width=0.45\linewidth]{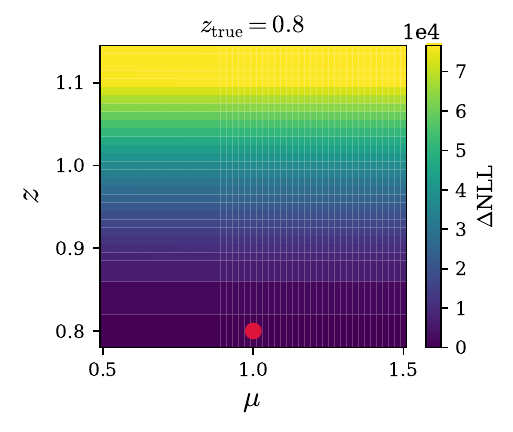}%
  \includegraphics[width=0.45\linewidth]{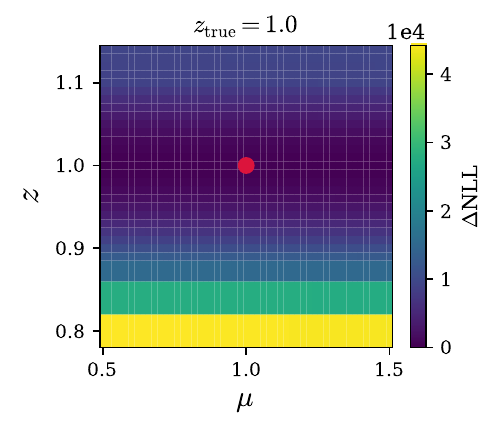}
  \includegraphics[width=0.45\linewidth]{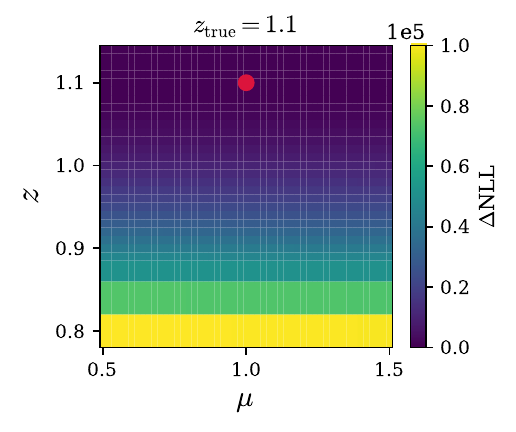}
  \caption{
      Value of the NLL of equation~\eqref{eq:nll}
      as a function of signal strength, $\mu$, and the nuisance parameter, $z$.
      The NLL values are obtained from scores obtained from the EBM
      assuming the observed data has a true value of nuisance parameter of
      $z_\mathrm{true} = 0.8$ (top-left), $z_\mathrm{true} = 1.0$ (top-right, nominal)
      and $z_\mathrm{true} = 1.1$ (bottom).
      The red dot signals  the minimum of the NLL\@.
      These figures are analogous to figure~11 in Ref.~\cite{Ghosh:2021roe},
      with the addition of $z_\mathrm{true} = 0.8$.
  }
  \label{fig:DNLL2D}
\end{figure}

Results for the optimisations performed using the loss functions of equations~\eqref{eq:betace} and~\eqref{eq:betasn} are shown in figure~\ref{fig:calibcompare}.
We also include the result we obtain using the Conditional MLP of Ref.~\cite{Ghosh:2021roe} (as ``Cond. MLP'', black dotted line). These results attemp to find the narrowest possible curve for the $\mu$ parameter. We see that the EBM with the cross-entropy optimised $\beta_\mathrm{ce}$ (red dashed line) performs worse than all the other cases, although very close to the result from the Conditional MLP. We also see that the EBM with $\beta_\mathrm{sn}$ optimised using signal templates (green solid line) provides the best result in all cases. In the case of $z = 0.8$ (left), below nominal, we see that all the lines remain very close.
However, for the nominal case (center), $z = 1.0$, we see that the cross-entropy optimised $\beta_\mathrm{ce}$ performs very closely to the Conditional MLP,
while the case of $\beta_\mathrm{sn}$, optimised with signal scores, has a clear separation from the other two. The case of $z = 1.1$ (right), above nominal,
again has the curves very close together, although with the case of the optimised $\beta_\mathrm{sn}$ having more separation than for the $z = 0.8$ case.
The values for the NLL as a function of $\mu$ and $z$ using our EBM are shown in figure~\ref{fig:DNLL2D}.
These figures are obtained from probabilities using $\beta_\mathrm{sn}$
and are analogous to the ones shown in figure~11 of Ref.~\cite{Ghosh:2021roe}.

Note that, while the optimisation of the loss function of equation~(\ref{eq:betace}) could be considered standard for classification, the availability of energies from the EBM and probabilities from the Gibbs distribution open a landscape of possibilities. One may envision the use of any of the steps from energy function design, to calibration of probabilities, including loss function setup, to improve the final purpose of the trained model. In our case, we have employed the very simple idea of calibrating $\beta$ with signal templates with the intent of improving the estimation of $\mu$ when systematic effects are present. However, many different ideas could be formulated for further improvements, on the same estimation. Access to energies and their probabilistic information could also be employed for more advanced analysis in several dimensions, including unbinned likelihood analysis. Some of these options are left for future explorations of EBMs in HEP.

\section{Summary and Conclusions}
{{In this study, a probabilistic energy-based NN framework was developed and its performance was  evaluated for the purpose of achieving uncertainty-aware classification in HEP collider analyses. The motivation for this work stemmed from the limitations of conventional discriminative classifiers, which often produce over-confident predictions for OOD events due to their reliance on softmax normalisation and the assumption that training and inference data share the same distribution. To address this flaw, the proposed approach utilised an EBM that assigns an energy score to each event, reflecting its compatibility with the learned data manifold.
The model architecture maps input features into a latent space using a NN, with class-specific embeddings serving as reference points. The energy was defined as the squared distance between the latent representation and the class embedding. This design enabled the EBM to distinguish between in-distribution and outlier events by calibrating a threshold on the energy score, which was directly linked to a quantifiable false/positive rate.}}

{{Experiments on a synthetic binary classification toy dataset demonstrated the effectiveness of this approach. The EBM achieved high classification accuracy on clean samples and robust noise detection. The energy landscape naturally formed closed regions around the training data, thereby allowing for reliable identification of noise without explicit supervision. Furthermore, the tunable energy threshold provides a flexible mechanism for balancing signal efficiency and noise contamination, with each threshold corresponding to a well-defined statistical uncertainty.}}

When applied to MC simulated collider events, specifically,
SM Higgs boson decays to $\tau^+\tau^-$ pairs simulated by ATLAS for LHC searches,
one finds that EBM performs scoring and classification
with results on par with other approaches based on composite networks.
Within the framework presented here,
the NN remained a rather simple MLP,
easy to setup, train, and evaluate.
However, the use of energies and derived probabilities
allowed the formulation of techniques to control
the uncertainty with the purpose of improving
signal strength estimation.

In conclusion, the probabilistic EBM-based NN framework offers a robust and interpretable approach for HEP collider analyses in the presence of experimental systematic uncertainties. Although its classification performance does not substantially exceed that of the conditional MLP baseline, its main advantage lies in the explicit and continuous control of uncertainty through the energy-based formulation. In particular, the calibrated energy score provides an event-level measure that can be propagated into the analysis and used to regulate the trade-off between classification performance and systematic uncertainty. This additional degree of control is not directly available in a conventional discriminative classifier and is especially useful when dealing with distributional shifts and poorly modelled events. Overall, the proposed approach achieves performance comparable to existing methods while providing a complementary uncertainty-aware mechanism that can be exploited in realistic LHC analyses. Its deployment in the context of actual LHC analysis  will be explored in a forthcoming publication.

\section*{Acknowledgments}
AH  is  partially supported by the Science, Technology and  Innovation Funding Authority (STDF) under grant number 50806. PK is supported by Basic Science Research Program through the National Research Foundation of Korea (NRF) funded by the Ministry of Education, Science and
Technology, RS-2025-24803289, and by KIAS Individual Grant No. PG021403. 
SM is supported in part through the NExT Institute and the STFC Consolidated Grant ST/C004240/1.
The work of RR is supported by a KIAS Individual Grant (QP094601)
via the Quantum Universe Center at the Korea Institute for Advanced Study (KIAS). 
Finally, this work used computational resources supported by the Center for Advanced Computation at KIAS\@.
\bibliographystyle{JHEP}
\bibliography{biblo}

\end{document}